\documentclass{article}

\usepackage[preprint]{neurips_2025}

\usepackage{graphicx}

\usepackage[utf8]{inputenc} 
\usepackage[T1]{fontenc}    
\usepackage{hyperref}       
\usepackage{url}            
\usepackage{booktabs}       
\usepackage{amsfonts}       
\usepackage{nicefrac}       
\usepackage{microtype}      
\usepackage{xcolor}         
\usepackage{wrapfig} 

\title{Computing with Canonical Microcircuits}

\author{%
 PK Douglas\thanks{} \\
  Queen Square Institute of Neurology\\
  University College London, UK\\
  IACS, Santa Monica, CA, US \\
  \texttt{pkdouglas16@gmail.com} \\
}

\begin{document}

\maketitle

\begin{abstract}
  The human brain represents the only known example of general intelligence that naturally aligns with human values. On a mere 20-watt power budget, the brain achieves robust learning and adaptive decision-making in ways that continue to elude advanced AI systems. Inspired by the brain, we present a computational architecture based on canonical microcircuits (CMCs)—stereotyped patterns of neurons found ubiquitously throughout the cortex. We implement these circuits as neural ODEs comprising spiny stellate, inhibitory, and pyramidal “neurons,” forming an 8-dimensional dynamical system with biologically plausible recurrent connections. Our experiments show that even a single CMC node achieves 97.8\% accuracy on MNIST, while hierarchical configurations—with learnable inter-regional connectivity and recurrent connections—yield improved performance on more complex image benchmarks. Notably, our approach achieves competitive results using substantially fewer parameters than conventional deep learning models. Phase space analysis revealed distinct dynamical trajectories for different input classes, highlighting interpretable, emergent behaviors observed in biological systems. These findings suggest that neuromorphic computing approaches can improve both efficiency and interpretability in artificial neural networks, offering new directions for parameter-efficient architectures grounded in the computational principles of the human brain.
\end{abstract}

\section{Introduction}

The architecture of the human brain is somewhat surprising.  Neurons are not connected at random, but rather form stereotyped circuits repeated throughout the cortex [1][2] [3][4]. These canonical circuit motifs serve as fundamental computational units—analogous to 'inception modules' in artificial networks—that collectively enable learning, memory, and a diverse array of complex cognitive functions [4][5].

\begin{wrapfigure}{r}{0.5\textwidth}
  \centering
  \vspace{-4mm}  
  \includegraphics[width=0.44\textwidth]{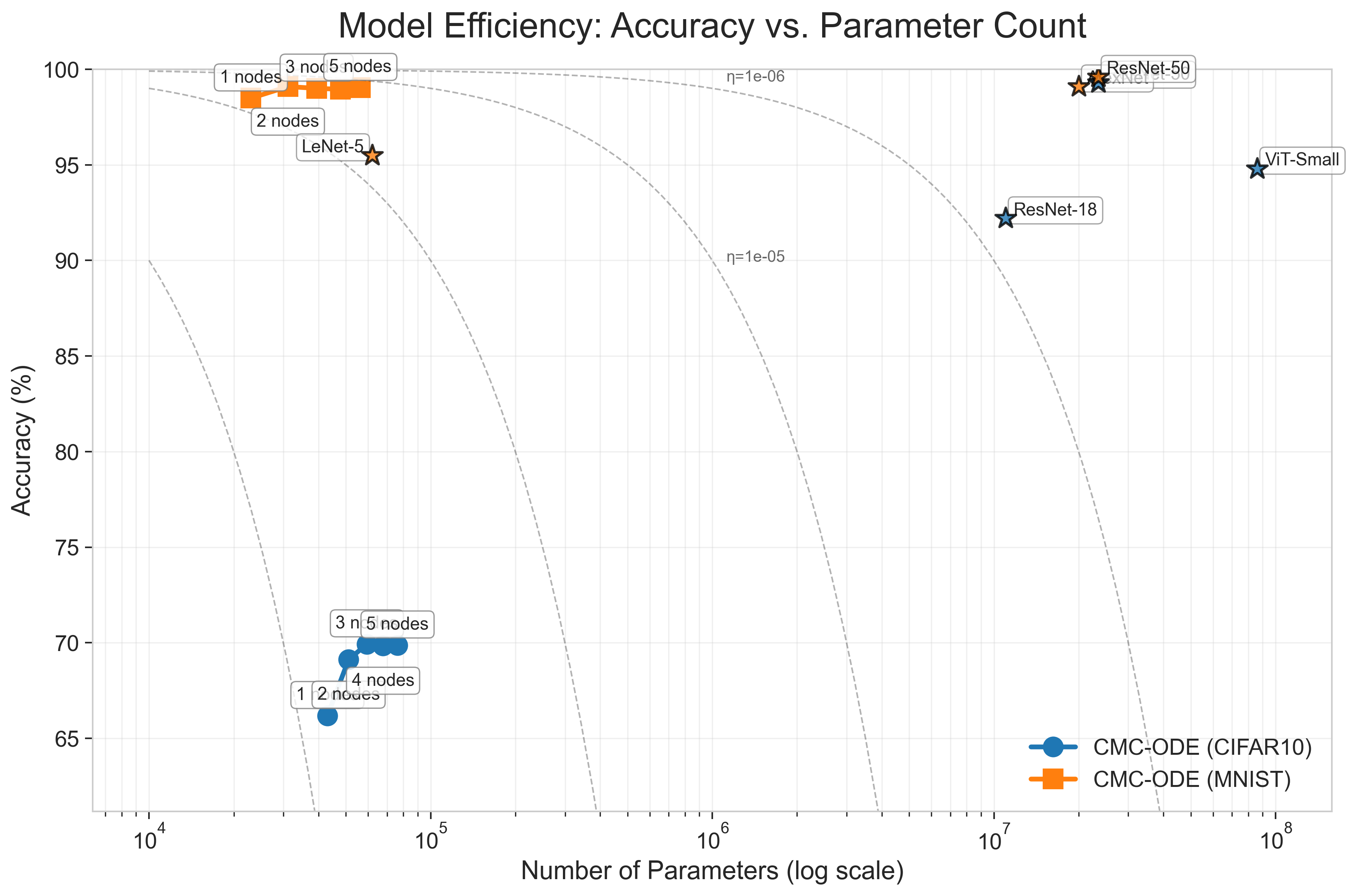}
  \vspace{-4mm}  
  \caption{\small Our brain-inspired model achieves competitive performance with orders of magnitude fewer parameters than conventional networks.}
  \vspace{-4mm}  
  \label{fig:teaser}
\end{wrapfigure}

Canonical microcircuits (CMCs) capture the population-level dynamics of functionally distinct neurons within a cortical column [6][7].  Each "neuron" in these models represents the collective behavior of a subpopulation of neurons sharing the same cell type, cortical layer, microcolumn membership, and functional tuning preference —an abstraction supported by converging empirical research across species e.g., [8].  Advanced techniques combining cellular spike recordings and optogenetics have revealed that columnar circuitry follows common patterns characterized by recurrent excitatory and inhibitory feedback loops.  Importantly, this pattern is observed not only in sensory and motor areas but also in association areas that subtend the highest levels of cognitive processing [4], establishing the microcolumn as a fundamental unit of information integration and functional specificity.

The computational power of these circuits arises not just from their individual dynamics but from their hierarchical embedding, small world network topology, fine-tuned input selectivity, and massive parallel interaction amongst approximately 10\textsuperscript{6} columns.  Research has demonstrated that differential weighting of cortical layers is crucial for working memory [9], attention [10], sensory and predictive processing [11], and altered states such as anesthesia [12].

AI has a long history of drawing inspiration from neuroscience spanning decades [13] [14]. However, while cross-pollination between these fields continues to generate excitement [15], such interdisciplinary exchange has become less common in recent years—potentially representing a significant missed opportunity.  Despite their fundamental importance in neuroscience, CMCs remain largely unexplored as computational architectures within the AI community. Contemporary deep learning models, while broadly inspired by neural principles, typically ignore the specific dynamics and recurrent connectivity patterns that characterize biological neural circuits. This divergence has potentially limited both the efficiency and interpretability of artificial neural networks.

\begin{figure}[t]
  \centering
\includegraphics[width=0.7\textwidth]{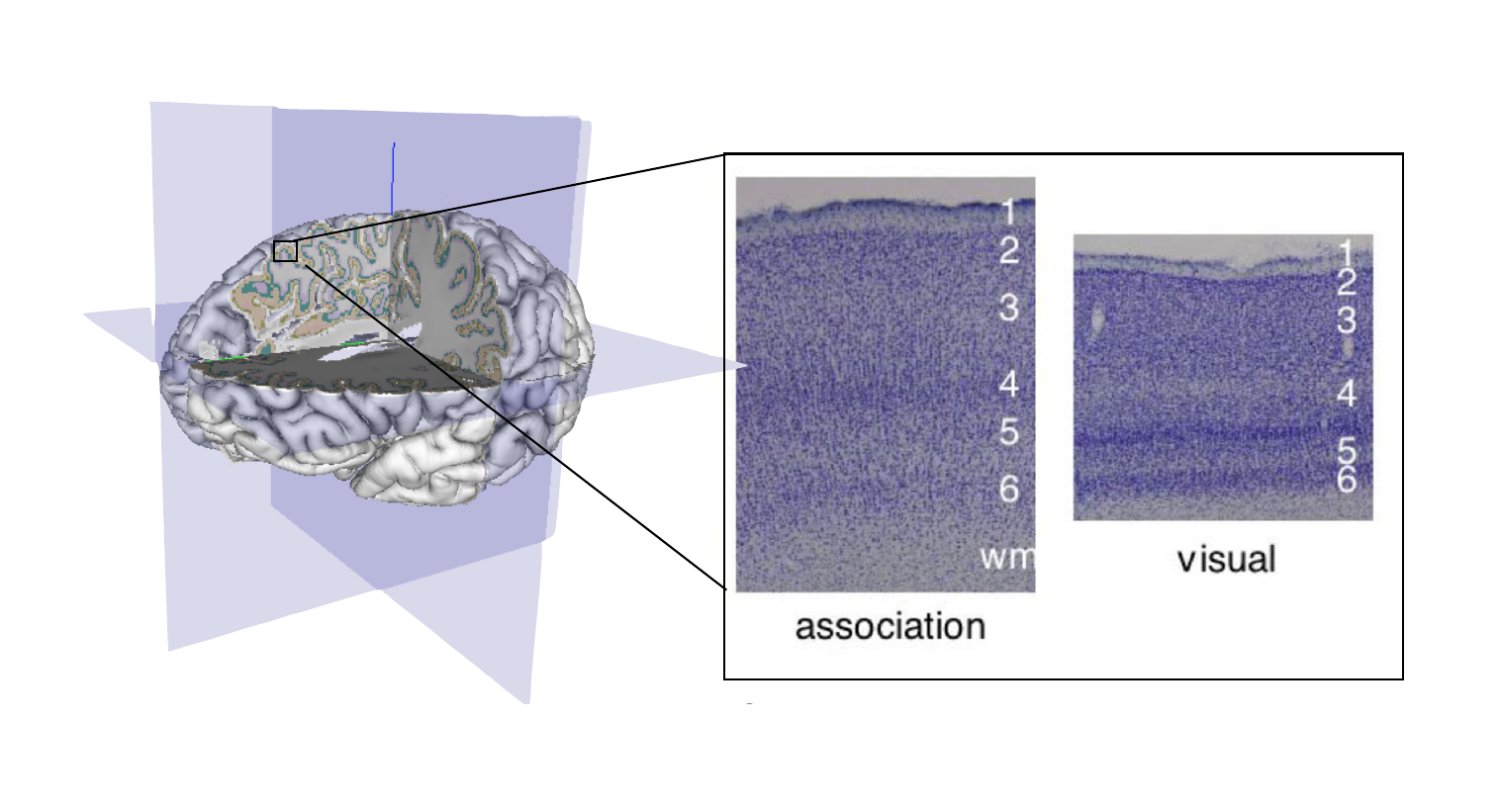}
\vspace{-4mm}  
  \caption{Laminar structure and repeated cytoarchitecture in the neocortex. Left: Cortical parcellation showing laminar-specific anatomy across cortical regions, based on cytoarchitectonic maps from the Julich Brain Atlas. Right: Nissl-stained histological sections illustrating consistent six-layered organization across distinct cortical areas. Nissl images adapted from publicly available data on brainmap.org.}
   \vspace{-4mm}  
  \label{fig:cmc_overview}
\end{figure}

In this paper, we introduce a novel computational framework that bridges neuroscience and artificial intelligence by translating CMCs into learnable neural ordinary differential equations (nODEs). Our approach generates dynamics of spiny stellate, inhibitory, and pyramidal neurons within an 8-dimensional dynamical system with biologically plausible recurrent connections. Unlike traditional deep learning architectures that abstract away biological details, we demonstrate that embracing neurobiological principles leads to remarkable computational efficiency (Figure 1). We extend this framework to hierarchical configurations that mirror brain information processing in the visual stream, establishing a new paradigm for neuromorphic computing. Through phase space analysis, we reveal interpretable dynamical properties that emerge naturally in our model, offering new insights into how neuromorphic computing principles might advance artificial neural network design. In summary, our contributions to the community include:
\begin{itemize}
\item A biologically-inspired nODE framework that integrates canonical microcircuit dynamics with retinal preprocessing, demonstrating how neuroscience principles can enhance deep learning architectures for vision tasks.

\item An empirical validation of scaling properties in flexible CMC models, showing how increasing the number of neural nodes (1-5) affects both accuracy and computational efficiency across multiple vision benchmarks.

\item State-of-the-art parameter efficiency demonstrating that CMC models with proper retinal preprocessing can match or exceed standard CNN performance while using orders of magnitude fewer parameters.

\item An open-source experimental framework with visualization tools for analyzing neural dynamics, phase portraits, and loss landscapes, enabling future research into interpretable CMC neural ODE models.
\end{itemize}

\section{Related Work}

\subsection{Canonical Microcircuits in Neuroscience}
Across the neocortex, a stereotyped pattern of connectivity, known as the canonical microcircuit (CMC), is found within vertically organized cortical columns (Figure 2). These columns, ranging from 50 to 500 $\mu$m in diameter, contain approximately 10\textsuperscript{2} to 10\textsuperscript{4} neurons and span the six cortical layers. Each microcolumn typically contains all the major neural phenotypes found in the neocortex, each of which localizes to a specific layer of the cortex during brain development (ontogeny) [3].  Although the cytoarchitecture is largely conserved across the cortex, each microcircuit becomes functionally specialized through learning and experience. This is conceptually similar to deep neural networks, where repeated architectural units can learn highly distinct representations depending on inputs and initial conditions [16].

Historically, computational models of canonical neural circuits have been developed in the computational neuroscience community, with the goal of reproducing empirical brain activity data. These models have evolved over the years to include more biologically realistic properties [1] [17] [7], such as nested feedback/feedforward loops [2], laminar specificity [6], and balanced excitation and inhibition [18]. These models have been used to model a range of phenomena to include visual recognition [19], syntax processing [20], neural oscillations [5], and attention [21].

Despite their rich theoretical development and empirical support, CMC models in their current neuroscientific instantiation have not, to our knowledge, been applied as computational building blocks for vision tasks in the AI community. Our work aims to fill this gap by reformulating CMCs as neural ODEs with learnable parameters, enabling their integration into scalable, trainable architectures while preserving their essential biological dynamics.

\subsection{Neural ODE Models}
Differential equations offer a natural and powerful framework for modeling systems that evolve continuously over time—a fundamental characteristic of biological systems.  The connection between neural networks and differential equations has been explored in a growing body of work, e.g., [22], [23] [24]. In particular, it has been shown that Residual Networks (ResNets) can be viewed as discrete approximations of continuous dynamical systems [22]. As the step size of the discretization approaches zero, this perspective naturally leads to neural ODEs (nODEs)—models that define hidden state evolution as the solution of a parameterized ODE. These models can be trained using traditional deep learning methods, and have shown strong performance on tasks such as modeling irregular time series [25] and continuous flows [26] with low computational burden. 

Neural Ordinary Differential Equations (Neural ODEs), first introduced by Chen et al. (2018), define hidden state evolution via parameterized differential equations, offering an alternative to traditional architectures with discrete layers. Importantly, nODEs are capable of capturing the recurrent, oscillatory behavior of neural populations—properties central to CMC models developed over decades in the neuroscience community. Neural ODEs thus offer a principled way to harness the computational power of deep learning with biologically inspired circuit motifs.

Recent work has extended the nODE framework in several directions. Latent ODEs have been used to model irregularly sampled time series in a generative latent space [27]. Augmented nODEs increase model capacity by expanding the dimensionality of the hidden state [28], while Neural Controlled Differential Equations (Neural CDEs) allow time-varying control signals to influence state trajectories [25]. Additional extensions include stochastic variational inference for latent stochastic differential equations [29], and Graph nODEs for solving continuous depth graph neural networks [30]. Neural memory ODEs (nmODEs) utilize a more biologically plausible learning algorithm without weight transport, for lower computational dimensions and greater efficiency [31].

\section{Modeling Canonical Microcircuits with Neural ODEs}

\subsubsection{Neural Dynamics}
We adopt a well-established formulation of canonical microcircuit dynamics widely used in computational neuroscience. Specifically, we implement the standard instantiation proposed by [6], which captures layer-specific cortical interactions among spiny stellate, pyramidal, and inhibitory neurons. Our goal is not to invent a new dynamical form, but rather to operationalize this biologically validated system as a scalable and trainable module for deep learning.  The CMC model comprises four primary neuron populations whose dynamics are governed by second-order differential equations as follows: 

\paragraph{Spiny stellate cells (granular layer):}
\begin{equation}
\ddot{v}_1 + 2\kappa_e\dot{v}_1 = \kappa_1 H_e(-\gamma_2\delta(v_4) + \gamma_1\delta(v_1) - \gamma_5\delta(v_2) + I)
\end{equation}

\paragraph{Inhibitory interneurons:}
\begin{equation}
\ddot{v}_2 + 2\kappa_i\dot{v}_2 + \kappa_i v_2 = \kappa_i H_i(-\gamma_{11}\delta(v_2) + \gamma_6\delta(v_3))
\end{equation}

\paragraph{Deep pyramidal cells (infragranular layer):}
\begin{equation}
\ddot{v}_3 + 2\kappa_e\dot{v}_3 + \kappa_e v_3 = \kappa_3 H_e(-\gamma_9\delta(v_2) - \gamma_{10}\delta(v_3))
\end{equation}

\paragraph{Superficial pyramidal cells (supragranular layer):}
\begin{equation}
\ddot{v}_4 + 2\kappa_e\dot{v}_4 + \kappa_e v_4 = \kappa_e H_e(-\gamma_8\delta(v_1) + \gamma_7\delta(v_4))
\end{equation}

Each second-order differential equation describes the post-synaptic membrane potential 
$v_i$ of neural population $i$, incorporating both intrinsic and extrinsic dynamics from connected populations, where: 
\begin{itemize}
\item $v_i$ : post-synaptic depolarizations of neural population $i$
\item $\kappa_{e}$ ,$\kappa_{i}$ : time constants for excitatory (e) and inhibitory populations (i)
\item $H_{e,i}(\cdot)$: average post-synaptic depolarizations 
\item $\delta(v_{j})$: nonlinear activation functions (e.g., sigmoid) mapping pre-synaptic to post-synaptic population dynamics; 
\item $\gamma_{v}$: Connectivity parameters from presynaptic to postsynaptic populations
\item I: exogenous input(e.g., sensory drive or thalamic input)
\end{itemize}

Replacing connectivity parameters $\gamma$, with a connectivity matrix over space and time enables one to generalize the neural mass model to a neural field model [7]. This dynamical system captures key motifs observed empirically in cortex, including recurrent excitation and laminar-specificity within a continuous-time framework compatible with neural ODE training (Figure 3). 

\begin{figure}[t]
  \centering
\includegraphics[width=0.72\textwidth]{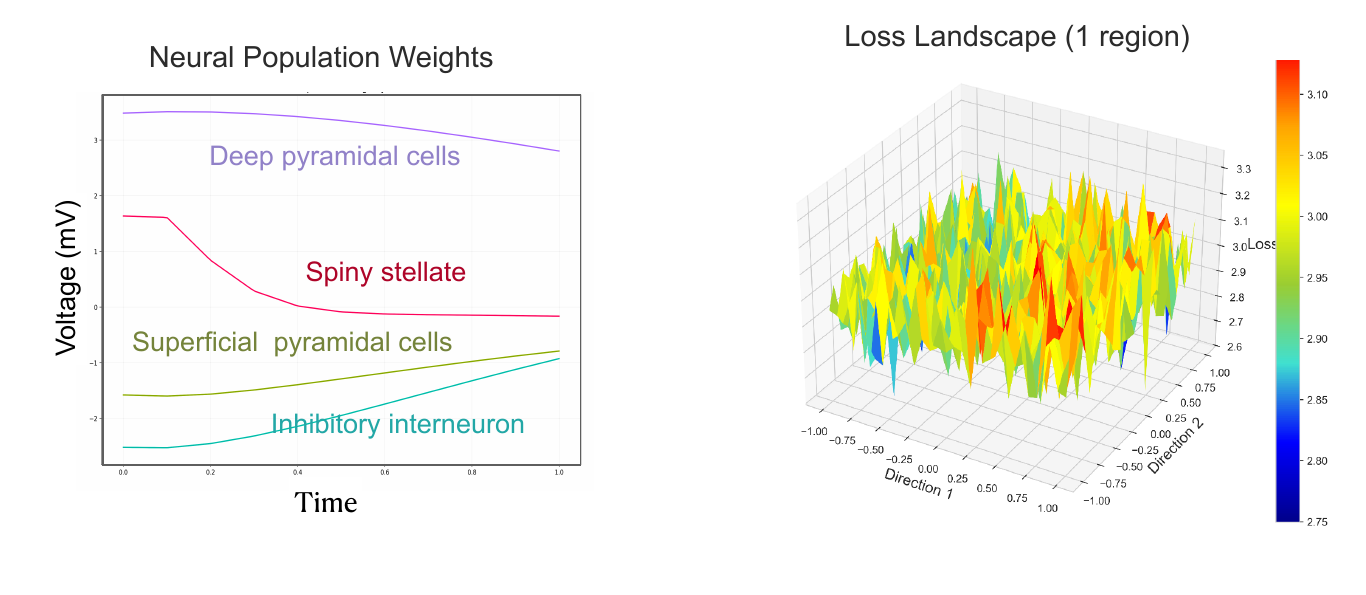}
  \caption{Single CMC node. Voltage dynamics (left) over time, and 3D loss landscape for a single-region CMC model on MNIST, revealing a rugged and highly sensitive optimization surface with multiple sharp local minima.  }
  \label{fig:new_mnist_loss}
\end{figure}

\section{Retinal Processing and Visual Hierarchy}
Visual processing in the brain proceeds along a hierarchical pathway from V1 to V5/MT, with each region specialized for progressively complex features [32]. This cascade has long inspired deep learning models, with parallels drawn between biological vision and convolutional networks like VGG [33]. We build on this tradition by embedding a canonical microcircuit (CMC) within a Neural ODE framework to model hierarchical dynamics in the visual system.

\subsection{Retinal Processing Layer}

The retina performs critical early-stage processing that fundamentally shapes visual information before it reaches cortical areas. We implemented a biologically-inspired retinal preprocessing module with the following components:
\begin{itemize}
\item Center-surround receptive fields using different kernel sizes (3×3 and 7×7) to simulate the multi-scale spatial processing of retinal ganglion cells
\item Parallel ON/OFF pathways modeled through separate convolutional channels with rectification, mimicking the complementary brightness detection circuits in the retina
\item Lateral inhibition implemented as a convolutional layer that enhances local contrast and edge detection
\item Adaptive gain control via learnable sigmoid-gated parameters that adjust channel sensitivity based on input statistics
\item Spatial downsampling through adaptive pooling to a standardized 14×14 feature map, reducing dimensionality while preserving essential visual structure
\end{itemize}

This retinal preprocessing transforms raw pixel inputs into a biologically plausible feature representation before projection into the CMC dynamics. The processed output is flattened and projected through a fully-connected layer to initialize the state variables of our downstream CMC nodes, which model higher cortical processing dynamics.  Our implementation balances biological fidelity with computational efficiency, creating a versatile front-end that can handle various visual datasets (MNIST, CIFAR, ImageNet) while maintaining reasonable parameter counts and training times.

\subsection{Hierarchical CMC (V1–V5)}
Following the retina, visual input is propagated through a series of CMCs, each representing a cortical visual area. These include: V1: Oriented edge and simple feature detection, V2: Boundary ownership, texture, and figure-ground segregation, V3: Shape and surface integration, V4: Mid-level feature and color representation, V5 (MT): Motion and depth estimation.  Each region is instantiated as an independent neural ODE module (Section 3.1) with internal excitatory/inhibitory dynamics and layer-specific recurrence. The depth and connectivity of the hierarchy are configurable, allowing us to explore both shallow (1–2 region) and deeper (5 region) CMC models.

\subsection{Inter-node Connectivity}
We implemented a flexible connectivity pattern between CMC nodes to model information flow across the cortical hierarchy including: 1) Inter-node connections are implemented via a learnable connectivity matrix, where each element represents the influence strength between pairs of nodes, 2) Connection strength is modulated through adaptive scaling via learnable parameters, with sigmoid gating to maintain stability during training, 
3) Input integration occurs by aggregating weighted activity from other nodes and incorporating it into each node's dynamic equations, and Influence mechanism is uniform across a node, taking the mean activity of the source node as input to the target node. 

Each node maintains the internal circuitry of the canonical microcircuit model with four neural populations (superficial pyramidal, spiny stellate, inhibitory interneurons, and deep pyramidal cells), but inter-node connectivity is implemented as a global modulatory influence rather than specific cell-type connections. This approach balances biological inspiration with computational simplicity, allowing the model to learn task-specific connectivity patterns through gradient-based optimization.

\section{Methods and Implementation}

\subsection{ODE Solvers and Optimization}

We implemented the CMC model in neural ODE format, resulting in an 8-dimensional dynamical system per region, where synaptic kernels were modeled using a configurable activation function (sigmoid, ReLU, or softplus), and all connectivity strengths $\gamma$ and synaptic time constants $\kappa$ were learnable.

For solving the ODEs, we initially implemented a 4th-order Runge-Kutta method, but found that a midpoint method or even Euler's method with appropriate step sizes offered a better balance between computational efficiency and numerical stability. The integration time span was set to [0,1] with 6 time points, using a step size of 0.15 for stability. To enhance convergence and numerical robustness, we applied several key optimizations: applying stability terms to prevent numerical instabilities, implementing tensor-based batched operations for parallel processing, and normalized state techniques to ensure convergence.  All source code is publicly available at: https://github.com/NeurotrustAI/CMC.

\subsection{Training Procedure}

The hierarchical architecture allows for a variable number of connected canonical microcircuits. Input data is projected through a convolutional network followed by a linear transformation that initializes the states of all brain regions. The regions are connected through both forward and backward pathways with learnable connection matrices, mimicking the top-down and bottom-up processing observed in the human brain. 

\begin{figure}[t]
  \centering
\includegraphics[width=0.7\textwidth]{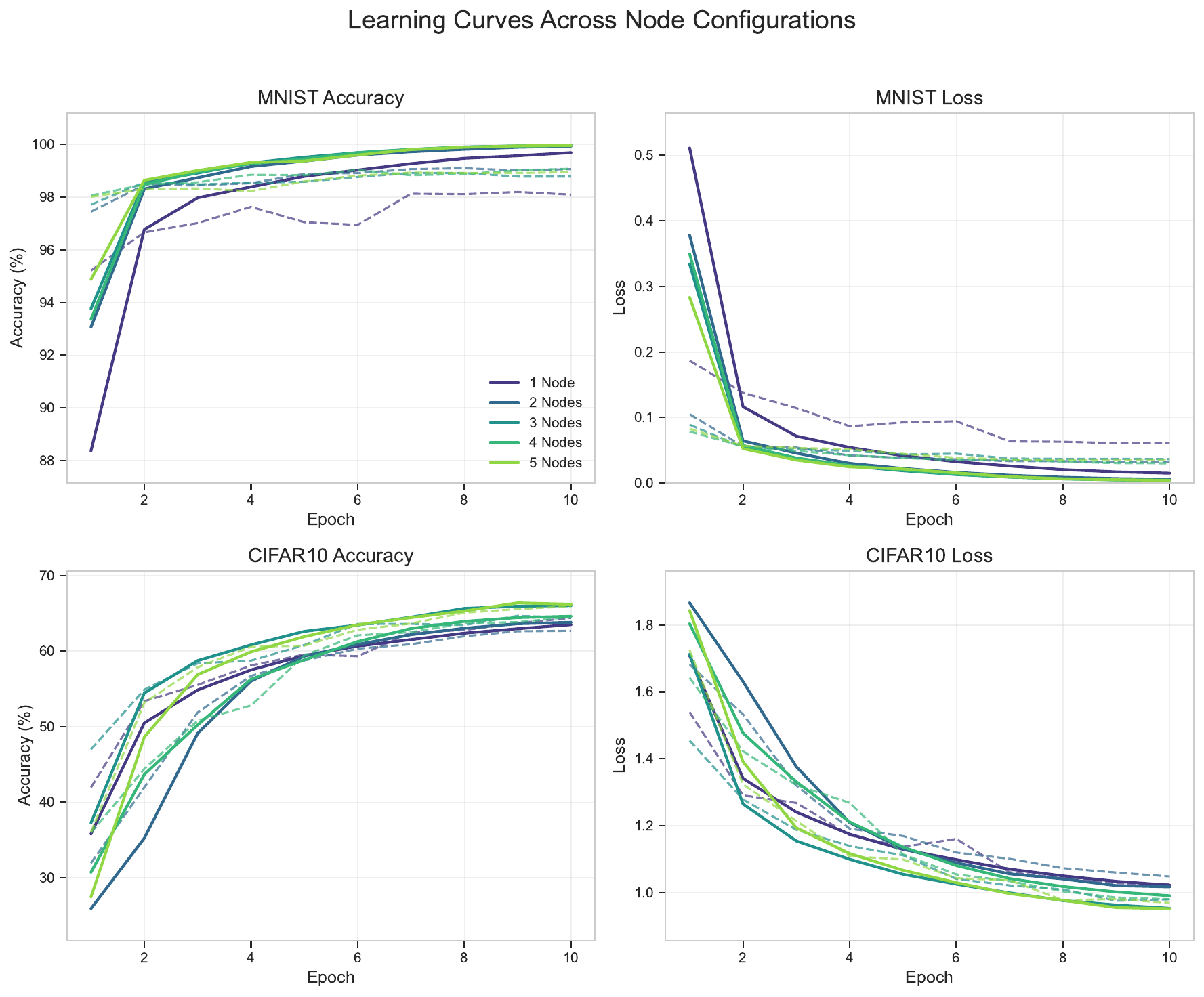}
  \caption{Top row: Accuracy and learning curves for CMC models with 1–5 regions on MNIST (top row), and CIFAR-10.}
  \label{fig:loss_landscape_CIFAR}
\end{figure}

Models were trained using the Adam optimizer with a cosine annealing learning rate schedule.  For MNIST, we achieved up to 99.8\% test accuracy with a 5-region model and 10 epochs. For CIFAR-10, accuracies reached 85.2\% with the same architecture. Training was performed with a batch size of 128, and the loss was computed using cross-entropy after mapping the final neural states through an output network (Figure 4). We implemented comprehensive tracking mechanisms to analyze the model's behavior including: phase trajectory capturing for specific MNIST digits, learning curve monitoring, loss landscape visualization, and neural voltage pattern analysis. 

\subsection{Computational Efficiency}

The ODE-based implementation introduces computational challenges due to the iterative nature of numerical integration. To mitigate this, we employed several efficiency strategies including: 1) unique and optimized solver parameters for each dataset, 2) connection scaling to maintain stable dynamics (0.1 scaling factor), 3) applied hardware acceleration where available (GPU/CUDA support), and 4) added fallback mechanisms to ensure robustness against numerical instabilities.

Parameter efficiency analysis revealed that our hierarchical model requires significantly fewer parameters than comparable deep learning architectures. A 5-region model contains approximately 150,000 parameters, compared to millions in CNNs like ResNet-18 or VGG networks with similar performance (Figure 1).

The accuracy-per-parameter metric showed that adding regions increases model capacity while maintaining parameter efficiency. The most efficient configuration was found to be 3 regions for MNIST and 4 regions for CIFAR-10, beyond which additional regions contributed diminishing returns in performance.

\section{Experiments}

We evaluated our neural ODE-based canonical microcircuit model across multiple experiments to assess its computational efficiency, representational capacity, and dynamic properties.

\begin{figure}[t]
  \centering
\includegraphics[width=0.72\textwidth]{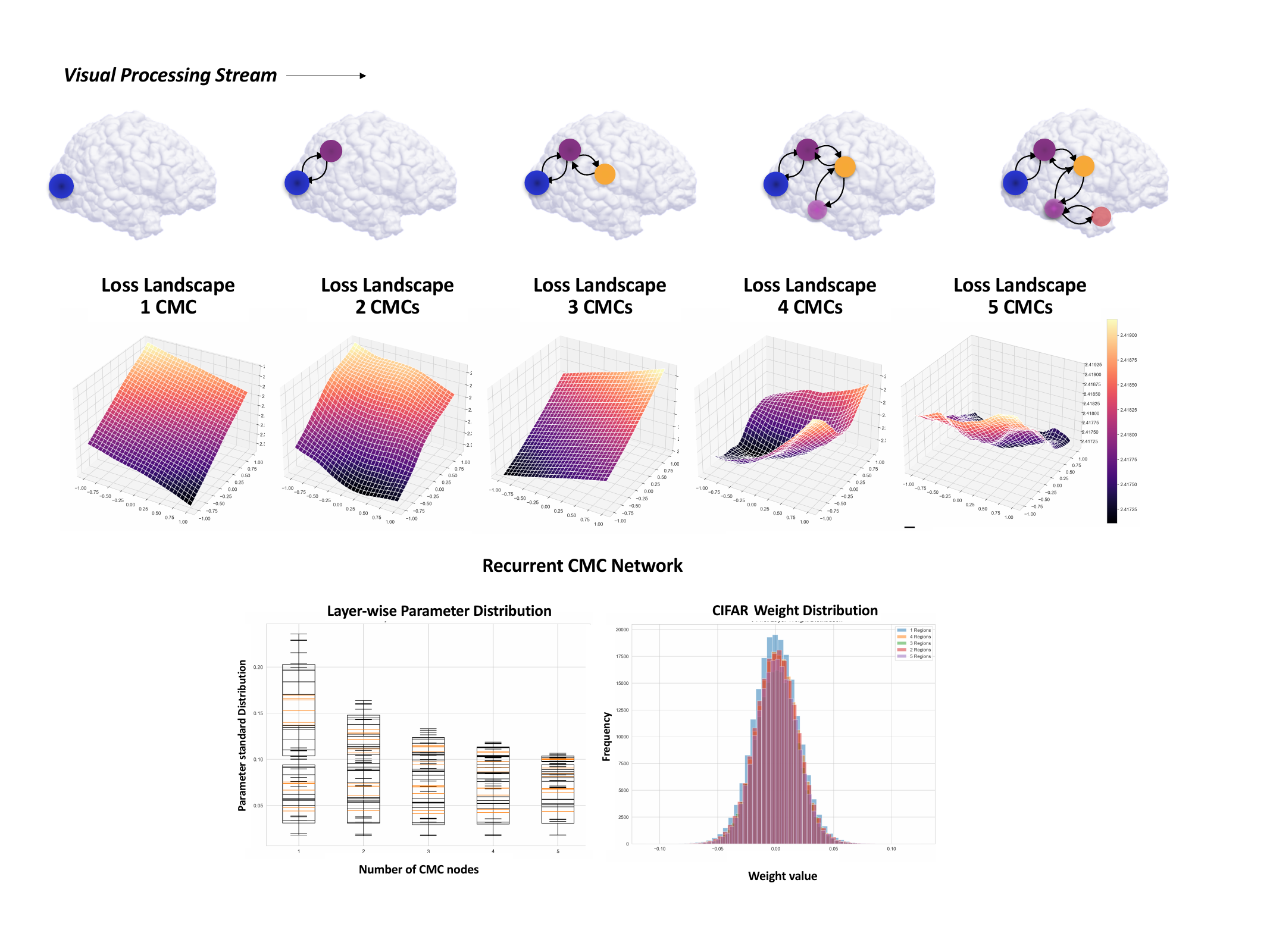}
  \caption{Effect of hierarchical depth on optimization and parameter structure in the CMC-nODE model (CIFAR-10).Top row: Loss landscapes for models with 1–5 CMC regions. Landscapes become increasingly complex with depth, yet remain smooth and convergent, suggesting robust gradient flow and high-capacity expressivity. Bottom row: Parameter distributions across CMC regions. Early regions exhibit broad variability, while deeper regions show tighter distributions, indicating progressive specialization and architectural regularization.  }
  \label{fig:loss_landscape_CIFAR}
\end{figure}

\subsection{Experimental Setup}

Datasets: We tested our model on standard computer vision benchmarks including MNIST, CIFAR-10, providing increasing levels of complexity for evaluation.
Model Configurations: We systematically varied the number of CMC nodes (1-5) to investigate how hierarchical depth affects performance. Each node implements the canonical microcircuit dynamics described in Section 3, with inter-node connectivity following the principles outlined in Section 4.3.
Training: All models were trained using Adam optimizer with a cosine annealing learning rate schedule. We ran multiple independent trials (n=3) for each configuration to ensure statistical reliability. For computational efficiency, we standardized batch sizes at 128 across all experiments.
Implementation: Models were implemented in PyTorch using the torchdiffeq package for ODE solving. The ODE solver used Euler integration with a step size of 0.05, which provided a good balance between computational efficiency and numerical stability.

\subsection{Classification Performance}

Figure 2 shows test accuracy across datasets as we increase the number of CMC nodes. For MNIST, performance improved substantially when scaling from 1 to 4 nodes (86.4\% → 95.2\%), with diminishing returns for additional nodes. For the more complex CIFAR-10 dataset, improvement continued through 5 nodes, suggesting that deeper hierarchical processing benefits more challenging visual tasks.

The performance scaling aligns with neuroscience findings that suggest higher visual areas in the brain encode increasingly complex features. In our model, each additional node enables more abstract representations of the input, similar to how visual cortical areas V1 through V5/MT process increasingly complex visual features.

\subsection{Parameter Efficiency}

We compared our model's parameter efficiency against standard deep learning architectures (Figure 1). Despite having fewer parameters than conventional CNNs and Transformers, our biologically-inspired models achieved competitive accuracy with orders of magnitude fewer parameters (1.3e+03× fewer than ViT-Small). A single CMC-ODE model with only 1 node remarkably acheived 86.5\% accuracy on MNIST.  The 4 node CMC-ODE reaches 99.2\% accuracy on MNIST with only ~150K parameters, compared to modified ResNet architectures that require millions of parameters for similar performance (Figure 1). This efficiency stems from the recurrent processing within nodes and the structured connectivity patterns between them, which reduce redundant parameters while maintaining expressive power.

\subsection{Dynamic Representations}

We visualized the trajectory of neural population activity during inference. Figure 5 shows phase portraits of superficial pyramidal cell activity for different input classes, revealing how the dynamics converge to class-specific attractors. For MNIST digits, we observed that 1. different digit classes form distinct trajectories in state space, 2. Early processing regions (nodes 1-2) show more variable dynamics, while later regions demonstrate more stable attractors, 3. The temporal evolution of neural activity shows characteristic transient responses followed by convergence.  These dynamics mirror key properties observed in biological neural recordings, where stimulus-specific neural trajectories evolve over time before stabilizing.

Additionally, We tracked the voltage evolution of different neural populations (e.g., spiny stellate cells) during training. Figure 4 illustrates how these neural dynamics progressively adapt as training proceeds. Initially, neural voltages exhibited random fluctuations with minimal class-specific patterns. As training progressed, voltage trajectories became increasingly structured and class-dependent. By the final epochs, distinct voltage patterns emerged for different input categories, demonstrating the model's ability to learn meaningful dynamical representations.

\section{Analysis, Limitations, and Future Directions}

\paragraph{Toward Greater Biological Plausibility}

While our results demonstrate the promise of using CMC dynamics within nODE frameworks for visual processing, several areas remain for further development and refinement.  Our current model uses backpropagation through time (BPTT) for training. Although effective, this method lacks biological realism. Future work will explore biologically plausible credit assignment mechanisms, including local Hebbian updates, predictive coding variants, and dendritic error integration schemes [34].  Additionally, we employed a configurable nonlinear activation (sigmoid, ReLU, softplus) to approximate neural firing rates. This approach, while tractable, does not fully capture the diversity of neuronal input-output functions observed in the cortex. Incorporating nonlinearities derived from biophysical neuron models—such as conductance-based transforms—may provide greater fidelity and functional richness.

\begin{figure}[t]
  \centering
\includegraphics[width=0.9\textwidth]{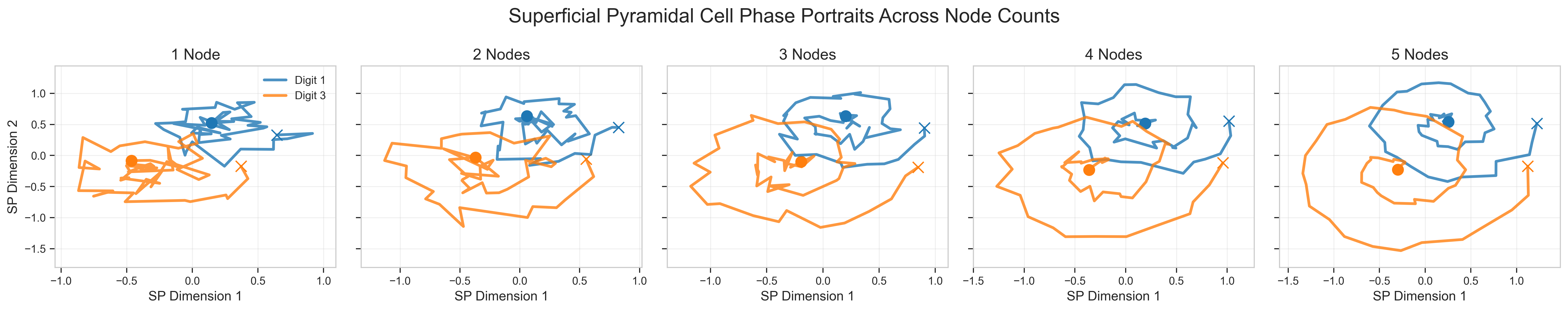}
  \caption{Phase portraits of superficial pyramidal cell activity for digits 1 (blue) and 3 (orange) across different CMC node counts (1-5). Panels show state space trajectory over training. As the number of nodes increases, trajectories become more distinct between digits, suggesting better class separation, and with later nodes (4-5) exhibiting more stable attractors. }
  \label{fig:phase}
\end{figure}

\paragraph{Architectural Extensions}
In this work, each cortical area (e.g., V1, V2) is modeled as a single CMC unit. However, biological brain regions contain multiple microcolumns operating concurrently. A natural extension of our model is to instantiate multiple CMCs per visual area, potentially enabling feature maps or retinotopic coverage across spatial receptive fields. We also omitted early subcortical processing stages such as the lateral geniculate nucleus (LGN) and optic chiasm decussation, which flip and reorganize visual fields before they reach V1. Including these components would provide a more faithful account of early visual preprocessing and could impact model symmetry, bias, and inter-hemispheric interaction.
Finally, while our model currently uses classical neural ODEs, more expressive formulations—such as augmented neural ODEs [28], neural controlled differential equations (Neural CDEs) [25], and hybrid spiking/ODE models—may improve representational dynamics and memory persistence.

\paragraph{Neuromodulatory Dynamics}

Biological circuits are shaped by neuromodulators like dopamine and norepinephrine, which dynamically modulate gain. For example, norepinephrine gates cortical processing by salience [35], while dopamine supports learning and reinforcement. Future work will extend our CMC-ODE framework to include neuromodulation, enabling simulation of global states like arousal and attention. These influences can be modeled as low-dimensional control signals on weights or gains.

\paragraph{Ablation Studies} We performed preliminary ablation studies to assess the role of various parameters and inter-regional feedback on model performance, see Supplementary Materials (Section A). These studies suggest that both architecture and neural population dynamics contribute substantially to overall performance.

\section{Discussion and Conclusions}
We introduced a biologically grounded neural architecture by translating the canonical microcircuit (CMC) model into a neural ODE framework. This approach captures structured cortical dynamics with high parameter efficiency, while maintaining competitive performance on standard vision tasks. Our results suggest that biologically inspired dynamical systems can serve as compact, interpretable modules in modern deep learning. Future extensions will explore greater biological realism, including neuromodulatory control, subcortical visual processing, and biologically plausible learning rules.

\begin{ack}

\end{ack}

\section*{References}

[1]	H. R. Wilson and J. D. Cowan, “Excitatory and Inhibitory Interactions in Localized Populations of Model Neurons,” Biophysical Journal, vol. 12, no. 1, pp. 1–24, Jan. 1972, doi: 10.1016/S0006-3495(72)86068-5.\\

[2]	R. J. Douglas and K. A. C. Martin, “NEURONAL CIRCUITS OF THE NEOCORTEX,” Annu. Rev. Neurosci., vol. 27, no. 1, pp. 419–451, Jul. 2004, doi: 10.1146/annurev.neuro.27.070203.144152.\\

[3]	V. Mountcastle, “The columnar organization of the neocortex,” Brain, vol. 120, no. 4, pp. 701–722, Apr. 1997, doi: 10.1093/brain/120.4.701.\\

[4]	K. D. Harris and G. M. G. Shepherd, “The neocortical circuit: themes and variations,” Nat Neurosci, vol. 18, no. 2, pp. 170–181, Feb. 2015, doi: 10.1038/nn.3917.\\

[5]	H. J. Luhmann, “Dynamics of neocortical networks: connectivity beyond the canonical microcircuit,” Pflugers Arch - Eur J Physiol, vol. 475, no. 9, pp. 1027–1033, Sep. 2023, doi: 10.1007/s00424-023-02830-y.\\

[6]	A. M. Bastos, W. M. Usrey, R. A. Adams, G. R. Mangun, P. Fries, and K. J. Friston, “Canonical Microcircuits for Predictive Coding,” Neuron, vol. 76, no. 4, pp. 695–711, Nov. 2012, doi: 10.1016/j.neuron.2012.10.038.\\

[7]	R. Moran, D. A. Pinotsis, and K. Friston, “Neural masses and fields in dynamic causal modeling,” Front. Comput. Neurosci., vol. 7, 2013, doi: 10.3389/fncom.2013.00057.\\

[8]	P. M. Goltstein, D. Laubender, T. Bonhoeffer, and M. Hübener, “A column-like organization for ocular dominance in mouse visual cortex,” Nat Commun, vol. 16, no. 1, p. 1926, Feb. 2025, doi: 10.1038/s41467-025-56780-3.\\

[9]	T. Van Kerkoerle, M. W. Self, and P. R. Roelfsema, “Layer-specificity in the effects of attention and working memory on activity in primary visual cortex,” Nat Commun, vol. 8, no. 1, p. 13804, Jan. 2017, doi: 10.1038/ncomms13804.\\

[10]	X. Wang, A. S. Nandy, and M. P. Jadi, “Laminar compartmentalization of attention modulation in area V4 aligns with the demands of visual processing hierarchy in the cortex,” Sci Rep, vol. 13, no. 1, p. 19558, Nov. 2023, doi: 10.1038/s41598-023-46722-8.\\

[11]	H. Adesnik and A. Naka, “Cracking the Function of Layers in the Sensory Cortex,” Neuron, vol. 100, no. 5, pp. 1028–1043, Dec. 2018, doi: 10.1016/j.neuron.2018.10.032.\\

[12]	A. M. Bastos et al., “Neural effects of propofol-induced unconsciousness and its reversal using thalamic stimulation,” eLife, vol. 10, p. e60824, Apr. 2021, doi: 10.7554/eLife.60824.\\

[13]	W. S. McCulloch and W. Pitts, “A logical calculus of the ideas immanent in nervous activity,” Bulletin of Mathematical Biophysics, vol. 5, no. 4, pp. 115–133, Dec. 1943, doi: 10.1007/BF02478259.\\

[14]	A. Zador et al., “Catalyzing next-generation Artificial Intelligence through NeuroAI,” Nat Commun, vol. 14, no. 1, p. 1597, Mar. 2023, doi: 10.1038/s41467-023-37180-x.\\

[15]	N. Kriegeskorte and P. K. Douglas, “Cognitive computational neuroscience,” Nat Neurosci, vol. 21, no. 9, pp. 1148–1160, Sep. 2018, doi: 10.1038/s41593-018-0210-5.\\

[16]	L. Wang et al., “Towards Understanding Learning Representations: To What Extent Do Different Neural Networks Learn the Same Representation,” Nov. 28, 2018, arXiv: arXiv:1810.11750. Accessed: May 19, 2022. [Online]. Available: http://arxiv.org/abs/1810.11750\\

[17]	R. J. Douglas and K. A. Martin, “A functional microcircuit for cat visual cortex.,” The Journal of Physiology, vol. 440, no. 1, pp. 735–769, Aug. 1991, doi: 10.1113/jphysiol.1991.sp018733.\\

[18]	S. Denève and C. K. Machens, “Efficient codes and balanced networks,” Nat Neurosci, vol. 19, no. 3, pp. 375–382, Mar. 2016, doi: 10.1038/nn.4243.\\

[19]	J. R. Gilbert and R. J. Moran, “Inputs to prefrontal cortex support visual recognition in the aging brain,” Sci Rep, vol. 6, no. 1, p. 31943, Aug. 2016, doi: 10.1038/srep31943.\\

[20]	T. Kunze, A. D. H. Peterson, J. Haueisen, and T. R. Knösche, “A model of individualized canonical microcircuits supporting cognitive operations,” PLoS ONE, vol. 12, no. 12, p. e0188003, Dec. 2017, doi: 10.1371/journal.pone.0188003.\\

[21]	F. Beuth and F. H. Hamker, “A mechanistic cortical microcircuit of attention for amplification, normalization and suppression,” Vision Research, vol. 116, pp. 241–257, Nov. 2015, doi: 10.1016/j.visres.2015.04.004.\\

[22]	R. T. Q. Chen, Y. Rubanova, J. Bettencourt, and D. K. Duvenaud, “Neural Ordinary Differential Equations,” in Advances in Neural Information Processing Systems, S. Bengio, H. Wallach, H. Larochelle, K. Grauman, N. Cesa-Bianchi, and R. Garnett, Eds., Curran Associates, Inc., 2018. \\

[23]	E. Weinan, “A Proposal on Machine Learning via Dynamical Systems,” Commun. Math. Stat., vol. 5, no. 1, pp. 1–11, Mar. 2017, doi: 10.1007/s40304-017-0103-z.\\

[24]	E. Haber and L. Ruthotto, “Stable Architectures for Deep Neural Networks,” 2017, doi: 10.48550/ARXIV.1705.03341.\\

[25]	P. Kidger, J. Morrill, J. Foster, and T. Lyons, “Neural Controlled Differential Equations for Irregular Time Series,” 2020, arXiv. doi: 10.48550/ARXIV.2005.08926.\\

[26]	W. Grathwohl, R. T. Q. Chen, J. Bettencourt, I. Sutskever, and D. Duvenaud, “FFJORD: Free-form Continuous Dynamics for Scalable Reversible Generative Models,” 2018, arXiv. doi: 10.48550/ARXIV.1810.01367.\\

[27]	Y. Rubanova, R. T. Q. Chen, and D. Duvenaud, “Latent ODEs for Irregularly-Sampled Time Series,” 2019, arXiv. doi: 10.48550/ARXIV.1907.03907.\\

[28]	E. Dupont, A. Doucet, and Y. W. Teh, “Augmented Neural ODEs,” 2019, arXiv. doi: 10.48550/ARXIV.1904.01681.\\

[29]	X. Li, T.-K. L. Wong, R. T. Q. Chen, and D. Duvenaud, “Scalable Gradients for Stochastic Differential Equations,” 2020, arXiv. doi: 10.48550/ARXIV.2001.01328.\\

[30]	M. Poli, S. Massaroli, J. Park, A. Yamashita, H. Asama, and J. Park, “Graph Neural Ordinary Differential Equations,” 2019, arXiv. doi: 10.48550/ARXIV.1911.07532.\\

[31]	X. Xu, H. Luo, Z. Yi, and H. Zhang, “A Forward Learning Algorithm for Neural Memory Ordinary Differential Equations,” Int. J. Neur. Syst., vol. 34, no. 09, p. 2450048, Sep. 2024, doi: 10.1142/S0129065724500485.\\

[32]	N. Kanwisher, “Functional specificity in the human brain: A window into the functional architecture of the mind,” Proc. Natl. Acad. Sci. U.S.A., vol. 107, no. 25, pp. 11163–11170, Jun. 2010, doi: 10.1073/pnas.1005062107.\\

[33]	N. Kriegeskorte, “Deep Neural Networks: A New Framework for Modeling Biological Vision and Brain Information Processing,” Annu. Rev. Vis. Sci., vol. 1, no. 1, pp. 417–446, Nov. 2015, doi: 10.1146/annurev-vision-082114-035447.\\

[34]	B. A. Richards et al., “A deep learning framework for neuroscience,” Nat Neurosci, vol. 22, no. 11, pp. 1761–1770, Nov. 2019, doi: 10.1038/s41593-019-0520-2.\\

[35]	A. C. Sales, K. J. Friston, M. W. Jones, A. E. Pickering, and R. J. Moran, “Locus Coeruleus tracking of prediction errors optimises cognitive flexibility: An Active Inference model,” PLoS Comput Biol, vol. 15, no. 1, p. e1006267, Jan. 2019, doi: 10.1371/journal.pcbi.1006267.



\newpage
\section*{NeurIPS Paper Checklist}

\begin{enumerate}

\item {\bf Claims}
    \item[] Question: Do the main claims made in the abstract and introduction accurately reflect the paper's contributions and scope?
    \item[] Answer: \answerYes{}
    \item[] Justification: We summarize the contributions clearly in the abstract and expand on them in Sections 1 and 3. 
    \item[] Guidelines: We followed all NeurIPS guidelines related to claims.

\item {\bf Limitations}
    \item[] Question: Does the paper discuss the limitations of the work performed by the authors?
    \item[] Answer:  \answerYes{}, 
    \item[] Justification: 
    YES. We created a specific section of the paper (Section 7) to directly address limitations related to the goals set forth in this paper.
    \item[] Guidelines: Guidelines: We followed all NeurIPS guidelines related to claims.

\item {\bf Theory assumptions and proofs}
    \item[] Question: For each theoretical result, does the paper provide the full set of assumptions and a complete (and correct) proof?
    \item[] Answer:  \answerNA{}.

\item {\bf Experimental result reproducibility}
    \item[] Question: Does the paper fully disclose all the information needed to reproduce the main experimental results of the paper to the extent that it affects the main claims and/or conclusions of the paper (regardless of whether the code and data are provided or not)?
    \item[] Answer: \answerYes{},
    \item[] Justification: Yes. We clearly describe our model architecture, training procedure, and experiments. Further, we have provided a Github link to our code repository that is publicly available and reproduces the work described herein.

\item {\bf Open access to data and code}
    \item[] Question: Does the paper provide open access to the data and code, with sufficient instructions to faithfully reproduce the main experimental results, as described in supplemental material?
    \item[] Answer: \answerYes{}
    \item[] Justification: Yes, a link to our publicly available and open source Github repository for this work is provided in the paper. 
    \item[] Guidelines: We followed all guidelines with respect to NeurIPS code.

\item {\bf Experimental setting/details}
    \item[] Question: Does the paper specify all the training and test details (e.g., data splits, hyperparameters, how they were chosen, type of optimizer, etc.) necessary to understand the results?
    \item[] Answer: \answerYes{}
    \item[] Justification: Yes, we clearly describe all parameters for training and testing, including the optimizer. All code is publicly accessible via the link provided. 
    \item[] Guidelines: We followed all NeurIPS guidelines for reporting. 

\item {\bf Experiment statistical significance}
    \item[] Question: Does the paper report error bars suitably and correctly defined or other appropriate information about the statistical significance of the experiments?
    \item[] Answer: \answerYes{}
    \item[] Justification: Yes, our figures show error bars or intervals for the main experimental figures where appropriate.  
    \item[] Guidelines: We followed all NeurIPS guidelines for reporting.

\item {\bf Experiments compute resources}
    \item[] Question: For each experiment, does the paper provide sufficient information on the computer resources (type of compute workers, memory, time of execution) needed to reproduce the experiments?
    \item[] Answer: \answerYes{}
    \item[] Justification: As described in the document, we used a CPU for these experiments. The code provided allows flexibility with respect to resources available, as described in the manuscript. 
    \item[] Guidelines: Guidelines followed.

\item {\bf Code of ethics}
    \item[] Question: Does the research conducted in the paper conform, in every respect, with the NeurIPS Code of Ethics \url{https://neurips.cc/public/EthicsGuidelines}?
    \item[] Answer: \answerYes{}
    \item[] Justification: Yes, we read the linked material to the code of ethics. 
    \item[] Guidelines: Yes all research was conducted to conform, in every respect, with the NeurIPS Code of Ethics .

\item {\bf Broader impacts}
    \item[] Question: Does the paper discuss both potential positive societal impacts and negative societal impacts of the work performed?
    \item[] Answer: \answerNA{}.
    \item[] Justification: We consider this to be foundational research, and not readily tied to any particular societal impact. 
    \item[] Guidelines: Guidelines followed.

\item {\bf Safeguards}
    \item[] Question: Does the paper describe safeguards that have been put in place for responsible release of data or models that have a high risk for misuse (e.g., pretrained language models, image generators, or scraped datasets)?
    \item[] Answer: \answerNA{}.
    \item[] Justification: No such risks
    \item[] Guidelines: Guidelines followed.

\item {\bf Licenses for existing assets}
    \item[] Question: Are the creators or original owners of assets (e.g., code, data, models), used in the paper, properly credited and are the license and terms of use explicitly mentioned and properly respected?
    \item[] Answer:\answerYes{}
    \item[] Justification: All works cited appropriately. 
    \item[] Guidelines: Guidelines followed.

\item {\bf New assets}
    \item[] Question: Are new assets introduced in the paper well documented and is the documentation provided alongside the assets?
    \item[] Answer: \answerNA{}.
    \item[] Justification: No assets used or obtained. 
    \item[] Guidelines: Guidelines followed. 
   
\item {\bf Crowdsourcing and research with human subjects}
    \item[] Question: For crowdsourcing experiments and research with human subjects, does the paper include the full text of instructions given to participants and screenshots, if applicable, as well as details about compensation (if any)? 
    \item[] Answer: \answerNA{}.
    \item[] Justification: Not used. 
    \item[] Guidelines: Followed.

\item {\bf Institutional review board (IRB) approvals or equivalent for research with human subjects}
    \item[] Question: Does the paper describe potential risks incurred by study participants, whether such risks were disclosed to the subjects, and whether Institutional Review Board (IRB) approvals (or an equivalent approval/review based on the requirements of your country or institution) were obtained?
    \item[] Answer: \answerNA{}.
    \item[] Justification: Human data not used. 
    \item[] Guidelines: Followed

\item {\bf Declaration of LLM usage}
    \item[] Question: Does the paper describe the usage of LLMs if it is an important, original, or non-standard component of the core methods in this research? Note that if the LLM is used only for writing, editing, or formatting purposes and does not impact the core methodology, scientific rigorousness, or originality of the research, declaration is not required.
    \item[] Answer:  \answerNA{}.
    \item[] Justification: Research not related to LLM development, etc. 
    \item[] Guidelines: Followed.

\end{enumerate}

\newpage
\section*{Supplementary Material: Ablation Studies and Architectural Variants}

To better understand the functional contributions of individual neuronal populations and architectural features within our Canonical Microcircuit (CMC)-based neural ODE framework, we conducted a series of ablation studies. These experiments were designed to assess how each component contributes to model performance, parameter efficiency, and the emergent dynamical properties observed in training.

Specifically, we performed systematic ablations of key neuronal populations within the CMC model—(1) deep pyramidal neurons, (2) spiny stellate cells, (3) superficial pyramidal neurons, and (4) inhibitory interneurons—as illustrated in Figure S7. For each ablation, we removed the corresponding state variables and updated the recurrent connectivity patterns accordingly. The modified neural ODEs for each variant are shown in Figure S8.

Beyond cell-type-specific ablations, we also tested a simplified model variant in which all inter-regional recurrent connections were removed, resulting in a purely feedforward composition of CMC nodes. These ablated architectures were evaluated across hierarchical CMC configurations using MNIST and CIFAR-10 benchmarks, with selected loss landscape results summarized in Figure S9.  The models with only feed foward connections show rugged loss surfaces with multiple local minima, suggesting a diminished capacity to generalize effectively to unseen data. 

To evaluate the functional contribution of each canonical microcircuit (CMC) cell type, we performed targeted ablation studies on the MNIST dataset using a hierarchical model spanning V1 to V3.  Figure S10 shows voltage dynamics for each active neuronal population, confirming successful ablations while preserving dynamics in other populations. Figure S11 summarizes the impact of different ablations on test accuracy across increasing hierarchy depth. Accuracy drops were modest but consistent across types (0.7–0.8\%).  In contrast, the feedforward-only variant (Figure S11, bottom-right panel) consistently underperformed its recurrent counterparts, highlighting the essential role of inter-regional feedback for robust inference. 

Performance degradation shows that spiny stellate cells were least critical (-0.8\% drop), compared to deep pyramidal cells being most critical with for a single (V1) CMC node. Interestingly, the original model was out-performed by each of the ablation tests for a single V1 node, suggesting that these ablations served to regularize the model with the MNIST data set.  Future testing may explore alternative models with increased regularization, and examine the extent to which this trend continues with more complex data sets. 
.

\begin{figure}[!b]
  \centering
\includegraphics[width=0.8\textwidth]{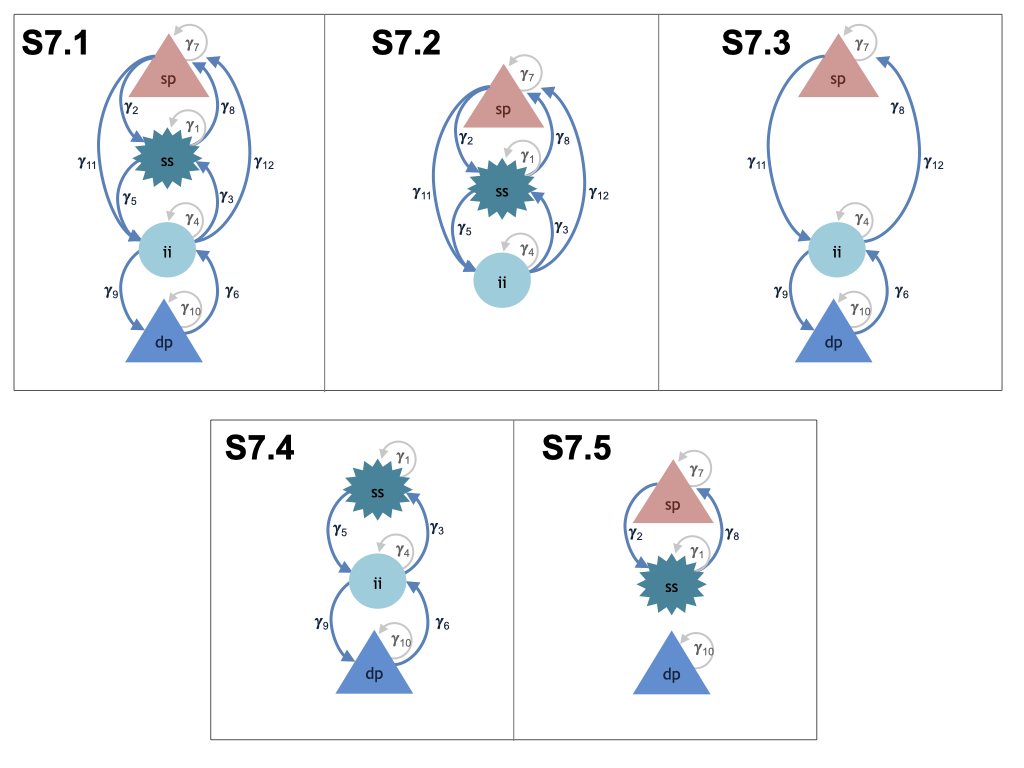}
  \caption{This figure depicts the full Canonical Microcircuit (CMC) model (top left, S7.1), alongside four ablation variants in which specific neuronal populations are removed: (S7.2) deep pyramidal neurons, (S7.3) spiny stellate cells, (S7.4) superficial pyramidal cells, and (S7.5) inhibitory interneurons. Each ablation selectively alters the flow of excitation and inhibition within the microcircuit, enabling analysis of the functional contribution of each component to the overall model dynamics and task performance.
 }
  \label{fig:ablation_schematic}
\end{figure}

\begin{figure}[t]
  \centering
\includegraphics[width=0.99\textwidth]{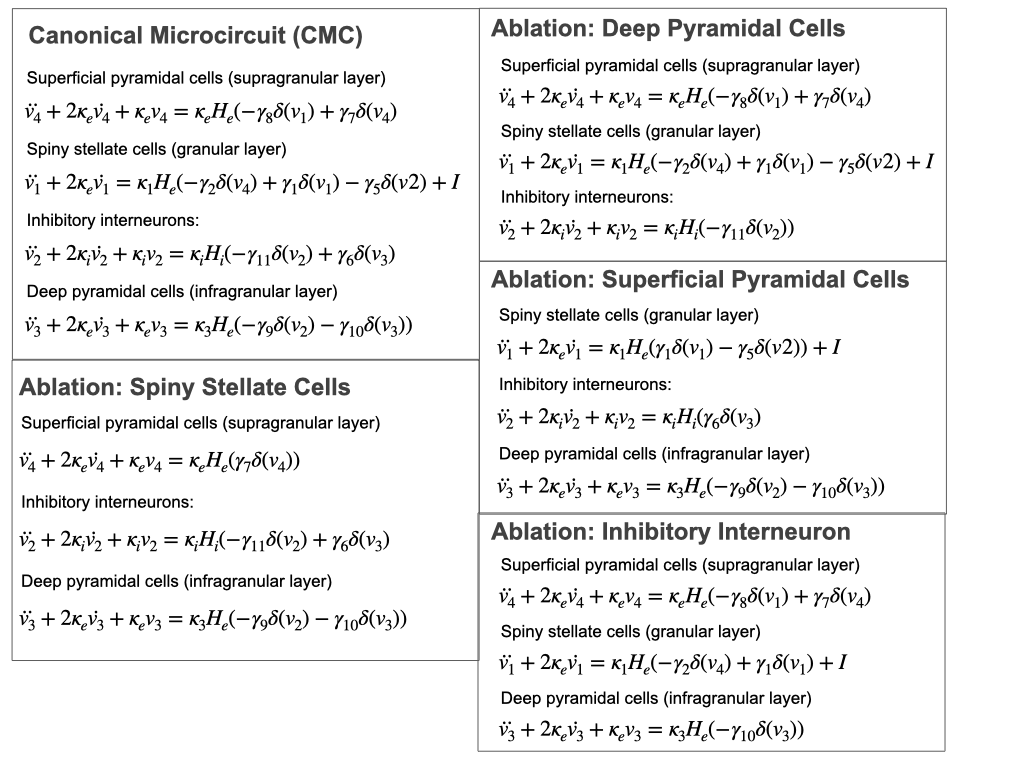}
  \caption{Neural ODE formulations for the original canonical microcircuit (CMC), and each neural population ablation in Figure S7. Each equation set reflects modifications to the original CMC dynamics, with the removal of specific populations resulting in the elimination of associated terms or feedback loops. 
 }
  \label{fig:ode_equations}
\end{figure}

\begin{figure}
  \centering
  \vspace{-3mm}
\includegraphics[width=0.8\textwidth]{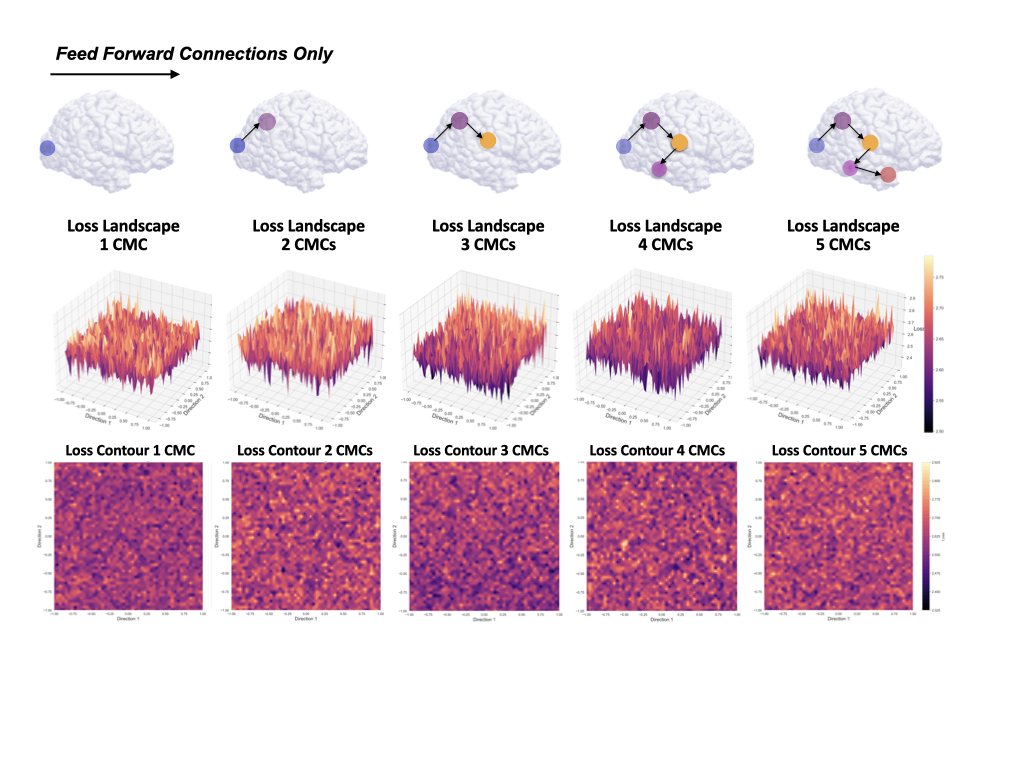}
  \vspace{-3mm}
  \caption{Loss landscapes for feedforward-only CMC models on CIFAR-10. Top row: cortical schematic for models with 1 to 5 Canonical Microcircuit (CMC) nodes, each composed in a purely feedforward hierarchy with no recurrent inter-regional connections. Center row: 3-d loss surfaces for each model, and Bottom row: corresponding 2D contour plots of the loss surface in parameter space. The landscapes appear rugged with deeper hierarchies, exhibiting sharp local minima and reduced smoothness compared to loss surfaces for the recurrent model. 
 }
  \label{fig:loss_CIFAR}
  \vspace{-4mm} 
\end{figure}

\begin{figure}
  \centering
  \vspace{-3mm}
\includegraphics[width=0.8\textwidth]{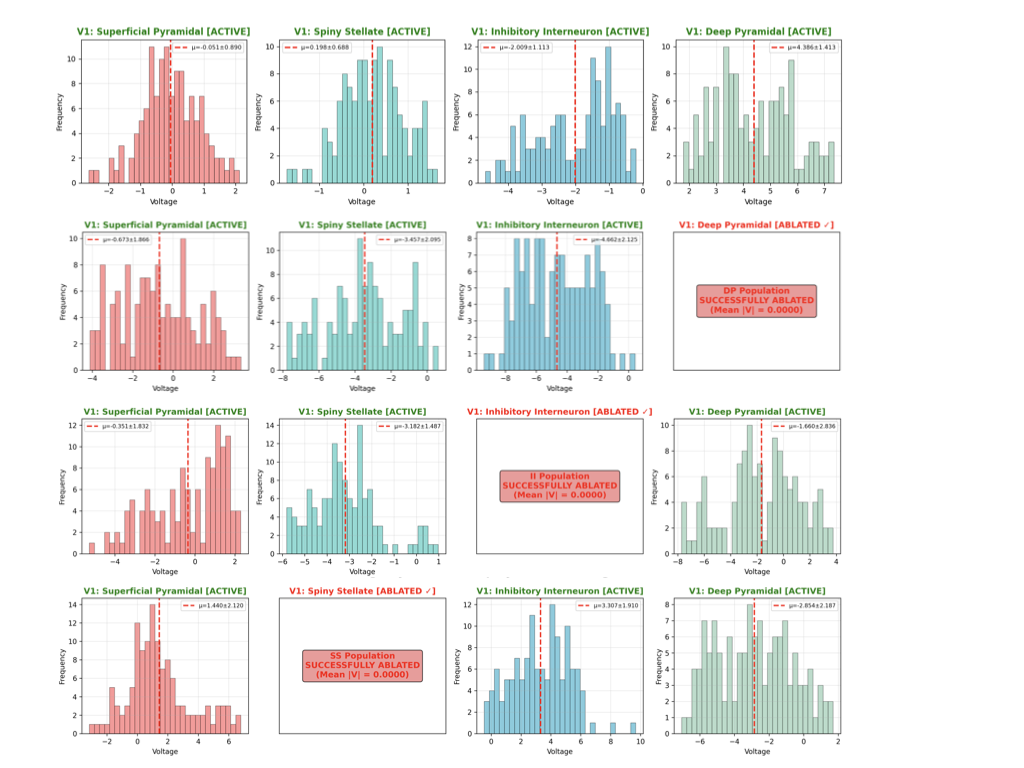}
  \vspace{-3mm}
  \caption{Voltage dynamics of neuronal populations in the V1 CMC node following cell-type-specific ablations (MNIST). Each panel shows the voltage distribution of a specific population (superficial pyramidal, spiny stellate, inhibitory interneuron, or deep pyramidal) under different ablation conditions. Red-highlighted boxes confirm successful ablation of the targeted population, with zero mean voltage activity. Non-ablated populations remain active, preserving distinct dynamics across variants. 
 }
  \label{fig:voltages}
  \vspace{-4mm} 
\end{figure}

\begin{figure}
  \centering
  \vspace{-3mm}
\includegraphics[width=0.8\textwidth]{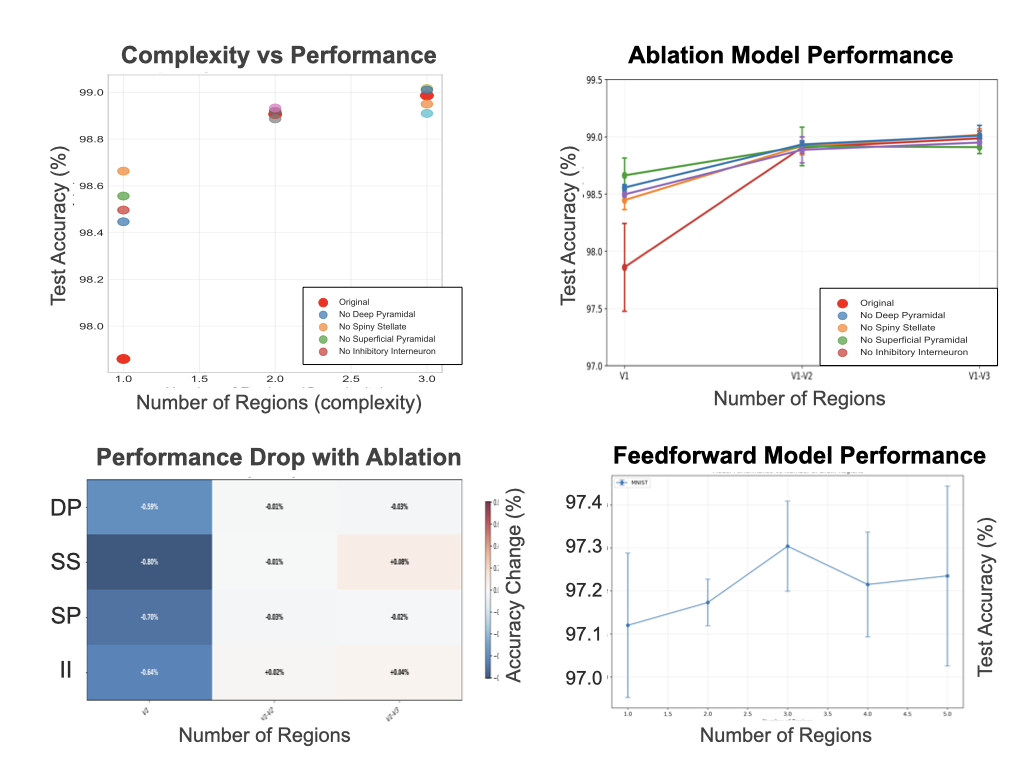}
  \vspace{-3mm}
  \caption{Hierarchical Ablation Analysis on MNIST. Top-left: Test accuracy as a function of model complexity (number of visual areas) for each ablation condition. Top-right: Ablation model performance across increasing hierarchical depth (V1 to V1–V3); accuracy generally improves with hierarchy regardless of ablation type. Bottom-left: Heatmap showing accuracy drop (\%) relative to the full model for each ablated population (DP = Deep Pyramidal, SS = Spiny Stellate, SP = Superficial Pyramidal, II = Inhibitory Interneuron). Bottom-right: Accuracy of the feedforward-only model (no inter-regional recurrence) across increasing regions; performance remains lower than recurrent variants, underscoring the benefit of recurrent dynamics for visual inference. 
 }
  \label{fig:s11}
  \vspace{-4mm} 
\end{figure}

\begin{figure}
  \centering
  \vspace{-3mm}
\includegraphics[width=0.8\textwidth]{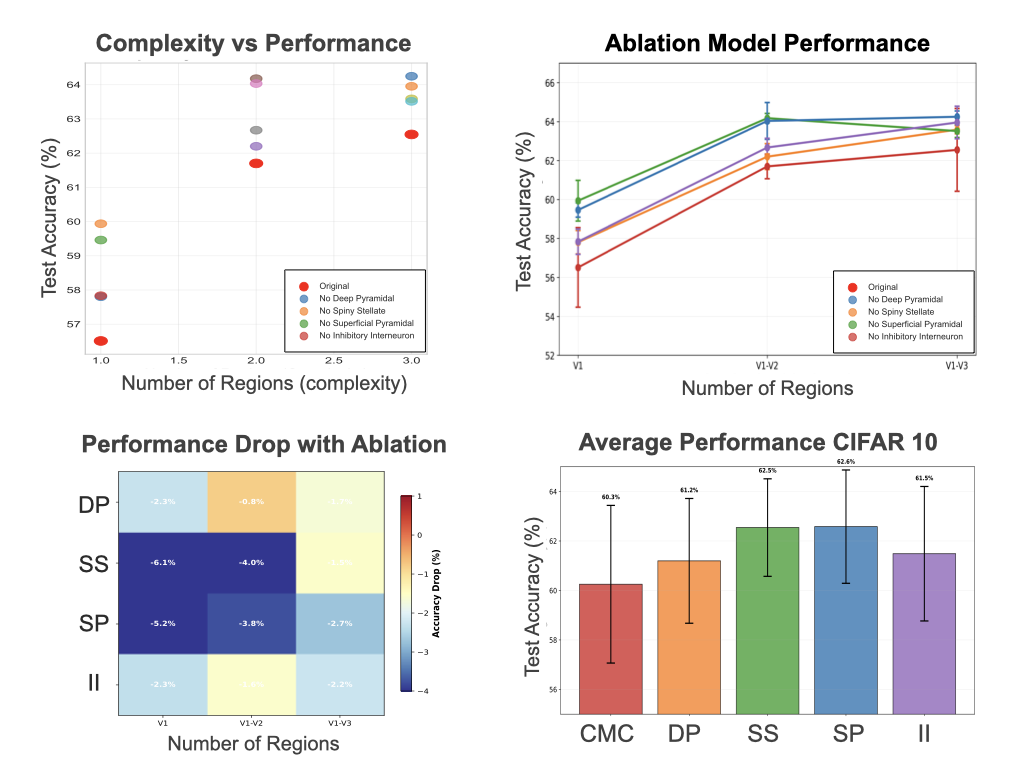}
  \vspace{-3mm}
  \caption{Hierarchical Ablation Analysis on CIFAR10. Top-left: Test accuracy as a function of model complexity (number of visual areas) for each ablation condition. Top-right: Ablation model performance across increasing hierarchical depth (V1 to V1–V3); accuracy generally improves with hierarchy regardless of ablation type. Bottom-left: Heatmap showing accuracy drop (\%) relative to the full model for each ablated population (DP = Deep Pyramidal, SS = Spiny Stellate, SP = Superficial Pyramidal, II = Inhibitory Interneuron). Bottom-right: Average performance for each ablation model. 
 }
  \label{fig:s12}
  \vspace{-4mm} 
\end{figure}

\end{document}